\documentclass[pre,superscriptaddress,twocolumn,footinbib]{revtex4-1}
\pdfoutput=1
\usepackage[colorlinks=true,urlcolor=blue]{hyperref}
\usepackage{amsfonts}
\usepackage{graphicx}
\usepackage{placeins}
\usepackage{amsmath}
\usepackage{xcolor}
\usepackage{color}
\usepackage{bm}

\definecolor{red}{rgb}{0.75,0,0}
\definecolor{blue}{rgb}{0,0,0.75}
\definecolor{green}{rgb}{0,0.5,0}

\begin{document}

\title{Defect order in active nematics on a curved surface}

\author{D. J. G. Pearce}
\affiliation{Dept. of Theoretical Physics, University of Geneva, Switzerland}
\affiliation{Dept. of Biochemistry, University of Geneva, Switzerland}
\begin{abstract}

We investigate the effects of extrinsic curvature on the turbulent behavior of a 2D active nematic confined to the surface of a cylinder. The surface of a cylinder has no intrinsic curvatrue and only extrinsic curvature. A nematic field reacts to the extrinsic curvature by trying to align with the lowest principle curvature, in this case parallel to the long axis of the cylinder. When nematics are sufficiently active, there is a proliferation of defects arising from a bend or splay instability depending on the nature of the active stress. The extrinsic curvature of the cylinder beaks the rotational symmetry of this process, implying that defects are created parallel or perpendicular to the cylinder depending on whether the active nematic is contractile or extensile.

\end{abstract}

\maketitle


Active matter describes a class of non-equilibrium materials in which stress is generated at the subunit scale; for example by converting some chemical energy source into directed motion \cite{Marchetti:2013,Ramaswamy:2010}. Active nematics are active materials which additionally have a broken rotation symmetry, resulting in nematic liquid crystalline phases \cite{Surrey:2001,Sanchez:2012,Doostmohammadi:2018,Needleman:2017,Marchetti:2013}. Active nematics have been observed as a naturally occurring phenomena in the motion of tissues \cite{Mueller:2019,Beng:2018} and bacteria \cite{Dunkel:2013,Wioland:2016,You:2018,Volfson:2008}, but can also be experimentally realized from reconstituted biological components; most commonly microtubule kinesin suspensions \cite{Surrey:2001,Sanchez:2012,Guillamat:2017,DeCamp:2015}. These active nematics show a huge range of phenomenological behaviors dependent on the level to which they are driven \cite{Giomi:2015,Giomi:2013,Giomi:2014}, the geometry of the enclosure \cite{Volfson:2008,Edwards:2009}, the properties of the substrate \cite{Pearce:2019,Guillamat:2017,Thijssena:2020}, the density \cite{You:2018} or the boundary conditions \cite{GiomiB:2014}. The novel features of active matter make it of particular interest in the design of new materials with non-trivial properties \cite{Thampi:2016,Souslov:2017}. This often requires that some control is exercised over the active matter which has been done previously through geometry and topology \cite{Souslov:2017,Thampi:2016,PearceB:2019,Pearce:2015,Ellis:2018,Keber:2014}.

The interplay between active nematics and substrate topology were first studied experimentally by Keber et al., who put active nematics on the surface of a vesicle \cite{Keber:2014}. This lead to a flurry of interesting results focusing on active nematics on curved surfaces \cite{Shankar:2017,Green:2017} and in particular spheres \cite{Keber:2014,Sknepnek:2015,Zhang:2016,Khoromskaia:2017,Mickelin:2018}. Using 3D printed water droplets it is possible to study the relationship between active nematics and surface Gaussian curvature on more complex surfaces \cite{Ellis:2018}, which has the effect of sorting topological defects into regions of like-sign Gaussian curvature \cite{Ellis:2018,PearceB:2019}. 

In this paper we study the effects of extrinsic curvature on topological defects within active nematics. This has been included in previous works, showing how it affects the lowest energy state for passive nematics \cite{Napoli:2016,Napoli:2018,Jesenek:2014}, the laminar flow regime in active nematics \cite{Napoli:2020}, and in generalized active nematics for curved surfaces \cite{PearceB:2019}. However its precise role in turbulent active nematics is not yet known. In order to isolate the effects of extrinsic curvature we study active nematics on the surface of a periodic cylinder; a developable surface with no Gaussian curvature. This also has the topological constraint of zero net topological charge within the nematic. We start by introducing the equations for an active nematic on a cylindrical surface, which requires only the addition of a single term to the free energy. We then go on to show how this affects the general behavior of an active nematic including the emergence of global nematic defect order. We show that this is an active effect by observing no such defect order for passive nematics on a cylinder. We then go on to propose a mechanism by which defect order is generated which we confirm by observing contractile active nematics.


Vectors on the surface of a cylinder can be written in cylindrical coordinates in the form $\bm{n} = n^\theta\bm{\hat{e}}_\theta + n^z\bm{\hat{e}}_z$, where $\bm{\hat{e}}_\theta$ and $\bm{\hat{e}}_z$ are basis vectors in the vertical and azimuthal directions, respectively. If we set our length-scale by the radius of the cylinder, this basis naturally becomes orthonormal, hence the surface metric can be written $g_{ij} = \delta_{ij}$. In addition, the basis vectors are independent of each other, hence the Christoffel symbols are all zero, $\Gamma_{ij}^{k}=0$. Finally, the surface of a cylinder is developable, therefore has no Gaussian curvature. This means that when writing the dynamical equations for a fluid on the surface of a cylinder, the intrinsic geometry of the surface has no effect on their form. Therefore we start from the generic form of the equations governing an in-compressible active nematic which are given by:

\begin{align}
\rho\partial_tv_i &= \partial_j\sigma^{(t)}_{ij} - \partial_jp\delta_{ij} - \mu v_j \label{eq:v}\\
[\partial_t + v_i\partial_i]Q_{ij} &= \lambda S u_{ij} + Q_{ik}\omega_{kj} - \omega_{ik}Q_{kj} + \gamma^{-1}H_{ij} \label{eq:Q}
\end{align}

Since we are in a cylindrical geometry $i$ and $j$ now run through $z$ and $\theta$, the vertical and azimuthal directions, respectively. Here $\rho$ is the density, $\sigma^{(t)}_{ij}$ is the total stress tensor and the $\mu$ is the friction coefficient. Since we are considering the in-compressible limit, $\partial_iv_i = 0$ and $\rho=1$ everywhere. $Q_{ij} = S(n_in_j - \delta_{ij}/2)$ is the nematic tensor, combining the nematic order parameter, $S$, and a unit vector pointing parallel to the average nematic director, $\bf{\hat{n}} = \textrm{cos}(\zeta)\bf{\hat{e}}_\theta + \textrm{sin}(\zeta)\bf{\hat{e}}_z$; here $\zeta$ is the angle between the average nematic orientation and the $\hat{\theta}$ direction. $\lambda$ is the flow alignment parameter coupling the evolution of the nematic tensor to the strain rate tensor, $u_{ij} = (\partial_iv_j + \partial_jv_i)/2$. The vorticity tensor is given by $\omega_{ij} = (\partial_iv_j-\partial_jv_i)/2$ and the molecular tensor is given by $H_{ij} = -\partial F/\partial Q_{ij}$ where $F$ is a Landau de Gennes free energy modified in order to account for the extrinsic curvature of the cylinder. 

\begin{equation}
F = \int dA \left[ \frac{K|\nabla Q|^2}{2} + \frac{K}{2\epsilon^2} trQ^2(trQ^2-1) + f_e\right]\label{eq:F}
\end{equation}

Where the parameter $\epsilon$ is a characteristic length which is proportional to the core defect radius and $K$ is the elastic constant associated with distortions in the director field. $f_e$ describes the energy cost associated with coupling the nematic tensor to the extrinsic curvature.

The extrinsic curvature tensor is given by $C_{ij} = -\bm{g}_i\cdot \partial_j \bm{\hat{N}}$; where $\bm{g}_i$ are the basis vectors and $\bm{\hat{N}}$ is the unit surface normal; on the surface of a cylinder this only has one non-zero component $C_{\theta\theta} = r$, the radius of the cylinder. The extrinsic curvature coupling energy is written $f_c = K_e(\bm{Q}.\bm{C}^2)$ which can be simplified to $f_c = -K_eQ_{\theta\theta}/r^2$ on a cylinder \cite{Jesenek:2014}. Since we have set our length scale by the radius of the cylinder, we consider only changes in $K_e$; this captures the same effect as a change in curvature. This energy describes the fact that the director embedded in the surface still has to follow the curvature of the surface as it is embedded in 3D space. This has the net effect of aligning the director with the central axis of the cylinder. If the director is aligned in the azimuthal direction, the director is curved within the embedding space since the surface of the cylinder is curved in that direction; this comes with an associated cost in elastic energy. However if the director is aligned in the vertical direction it experiences no such curvature, and hence has no energy penalty. It should be noted that if the fibers within the nematic had non-zero spontaneous curvature the effect would be reversed and the fibers would align with the cylinder at an angle which accommodates their curvature; a mechanism that has been proposed for the alignment of FtsZ fibers at bacterial division sites \cite{Bisson:2017,Caldas:2019,Gonzalez:2014}.

The total stress tensor ($\sigma^{(t)}$) is the sum of elastic stresses ($\sigma^{(e)}$), viscous stresses ($\sigma^{(v)}$), and the active stress generated by the molecular motors ($\sigma^{(a)}$) controlled by parameter $\alpha$.

\begin{align}
\sigma^{(e)}_{ij} &= -\lambda S H_{ij} + Q_{ik}H_{kj} - H_{ik}Q_{kj} \label{eq:sig_e}\\
\sigma^{(v)}_{ij} &= \nu u_{ij} \label{eq:sig_v}\\
\sigma^{(a)}_{ij} &= \alpha Q_{ij} \label{eq:sig_a}
\end{align}

Here $\alpha$ is the magnitude of the active stress and $\nu$ is the viscosity coefficient. The active stress is what separates this system from a passive liquid crystal. This describes the generation of stresses at the microscopic length scale which can either be extensile, $\alpha <0$, or contractile, $\alpha >0$. When the magnitude of the activity is sufficiently high, the additional active stress can overcome the elastic stresses in the system and there is a proliferation of topological defects and associated flow in a state known as ``active turbulence''. The lowest order defects in a 2D nematic system have a half integer charge associated with a half turn of the director as a line is traced around the core of the defect; this results in the characteristic $\pm 1/2$ nematic defects. The orientation of the director around the core is given by $\zeta = k(\phi - \psi) + \psi$ where $k$ is the charge of the defect, $\phi$ is the angle between a reference axis, in this case the $\hat{z}$ direction, and the position around the defect core \cite{Vromans:2016}. $\psi$ is the orientation of the defect which is polar in the case of $+1/2$ defects and triatic in the case of $-1/2$ defects. 

This set of equations can be easily derived from the generalised hydrodynamics of active nematics on curved surfaces equations presented previously \cite{PearceB:2019}. Equations \ref{eq:v}-\ref{eq:Q} are simulated with a periodic boundary in two dimensions recreating an active nematic on the surface of a periodic cylinder, see Supp. Movie. The model parameters are selected such that the system is in a state of active turbulence containing of the order of 100 defects, see S.I. for details.

\begin{figure}[]
\centering
\includegraphics[width=\columnwidth]{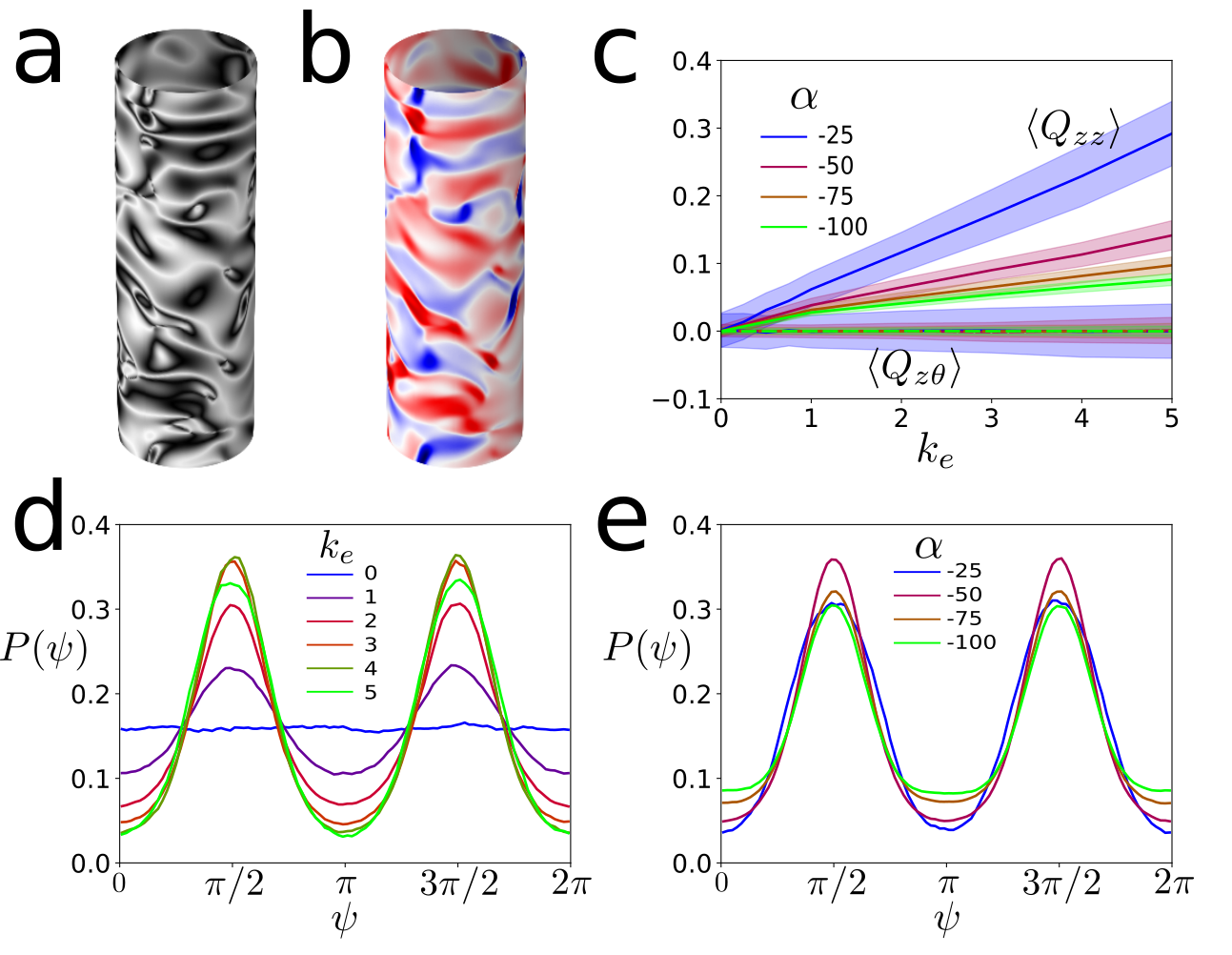}
\caption{\label{fig:active} (a-b) Snapshots of the Schlieren texture and vorticity field of a simulated extensile active nematic ($\alpha<0$) on the surface of a cylinder. (c) The nematic tensor shows increased average alignment with the $\hat{z}$ direction as $K_e$ is increased. (d) Extrinsic curvature coupling causes positive defects to align perpendicular to the $\hat{z}$ direction. (e) The activity has only a slight affect on the observed defect order.}
\end{figure}

As the extrinsic curvature coupling coefficient, $K_e$, is increased we see an increase in the average value of $Q_{zz}$, which implies a decrease in the average value of $Q_{\theta\theta}$, see Fig.~\ref{fig:active}a. This suggests that much of the active nematic shows net alignment with the cylinder, even in the active turbulence regime. This effect is strongest when the magnitude of the active stress is smallest. This is because the active stress acts to destabilize the ordered nematic state and introduce defects. Defects necessitate variations in the nematic director so it inherently becomes less ordered. The value of $Q_{z\theta}$ is uniformly low, meaning that there is no detected alignment in any other direction, see Fig.~\ref{fig:active}a. The defects in an active nematic on flat space show no global ordering \cite{Pearce:2019}, however when $K_e$ is increased we see an emergent nematic order in which the defects on average align themselves pointing parallel or anti-parallel to the azimuthal direction, see Fig.~\ref{fig:active}d. 

The density of defects in the active turbulent state depends on the activity and elastic coefficient; defects can be generated on the length-scale at which the active stress balances the elastic energy, \cite{Giomi:2013,Giomi:2015}. This gives a defect density that scales $\rho_D\sim\alpha/K$, hence increasing the activity increases the number of defects. The extrinsic curvature has a small effect on the number of defects (see SI) and there is only a slight effect of the activity on the ordering of defects on a cylinder, see Fig.~\ref{fig:active}e. 

\begin{figure}[]
\centering
\includegraphics[width=\columnwidth]{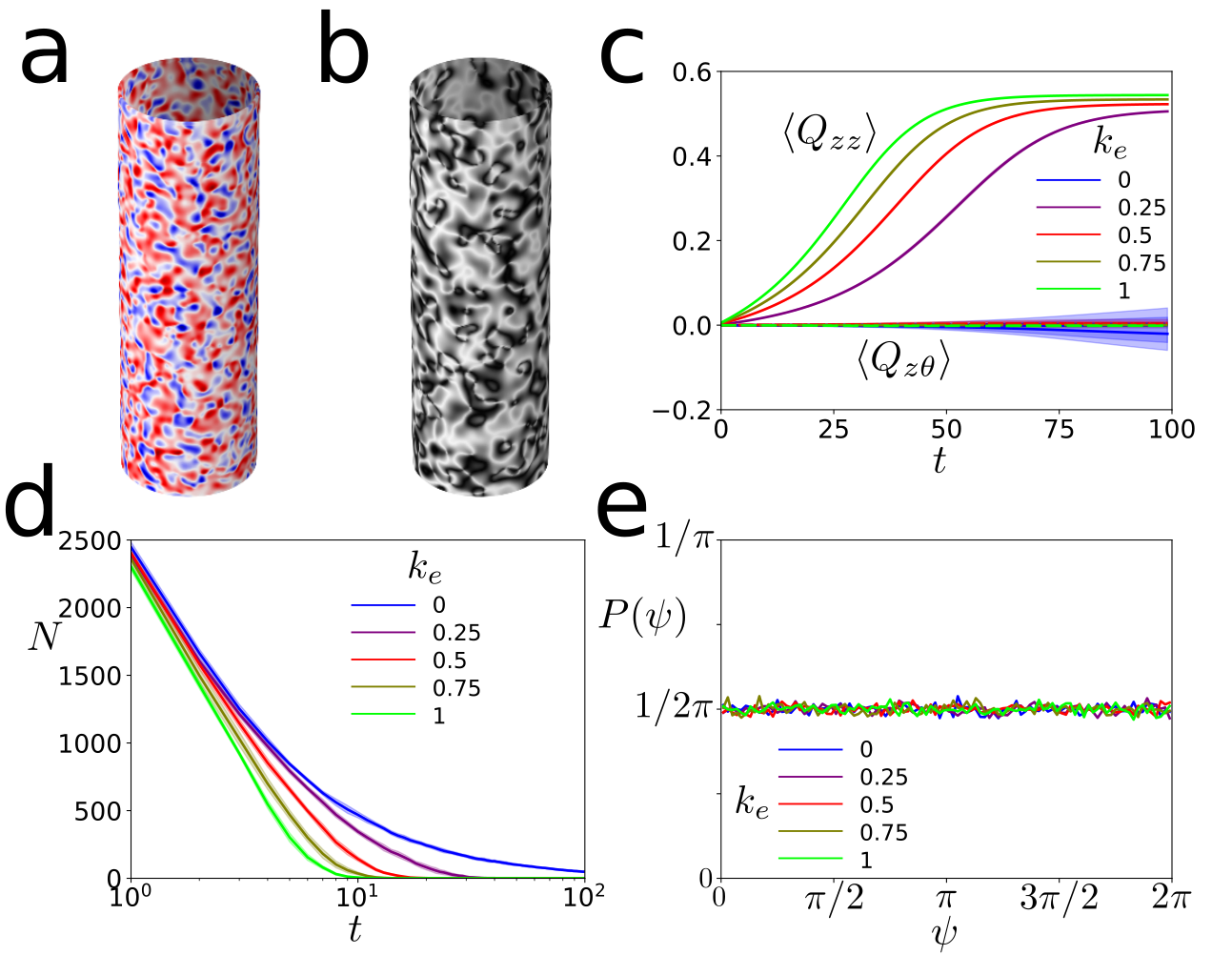}
\caption{\label{fig:passive}(a-b) Snapshots of the Schlieren texture and vorticity field of a simulated passive nematic on the surface of a cylinder. (c) The average nematic tensor aligns with the $\hat{z}$ direction when $K_e\neq0$. (d) The extrinsic curvature coupling causes an increased rate of decay for the defects. (e) No net alignment of defects relative to the cylinder is observed in passive nematic defects. }
\end{figure}

We can observe a passive nematic by removing the active stress, $\alpha=0$. In this case the system quickly approaches the minimum energy configuration with a uniform director aligned with the cylinder, as $K_e$ is increased the speed at which the system equilibrates is faster, see Fig.~\ref{fig:passive}a. The system is initialized with a randomized director which contains many defects which annihilate over the course of the simulation. This coarse-graining of the defects also processes faster when $K_e$ is increased, see Fig.~\ref{fig:passive}b. This is because the extrinsic curvature breaks the rotational symmetry of the system, hence there is no spontaneous symmetry breaking required. By repeating the coarse-graining process many times with random initial conditions it is possible to build up enough statistics to look for order within the defects. In the passive case we see no anisotropy in the orientations of the defects, see Fig.~\ref{fig:passive}c. 

The passive system is entirely driven by relaxation of the elastic energy. The elastic energy associated with a defect is described by the modified Landau de Gennes free energy given above, see Eq.~\ref{eq:F}. The only part of this energy which breaks rotational symmetry with respect to the cylinder is the extrinsic curvature energy $f_e$. Therefore we can write the elastic torque on a defect core to be $\tau = \partial_\psi \int_{\Delta^2}dA F = \partial_\psi \int_{\Delta^2}dA f_e$, where $\Delta^2$ is a small neighborhood around the core of a defect. By substituting in the expression for the director around the core of a defect it is easy to arrive at the following result. 

\begin{equation}
\tau = \partial_\psi \int_{\Delta^2}f_edA  = \int_{\delta r}\int_0^{2\pi}\rm{cos}(\psi + \phi)rd\phi dr = 0
\end{equation}

Hence there are no elastic torques on the defects which explains why we observe no orientational ordering of defects within the passive system. Therefore the ordering we observe in the active system must come from the active stress.

\begin{figure}[]
\centering
\includegraphics[width=\columnwidth]{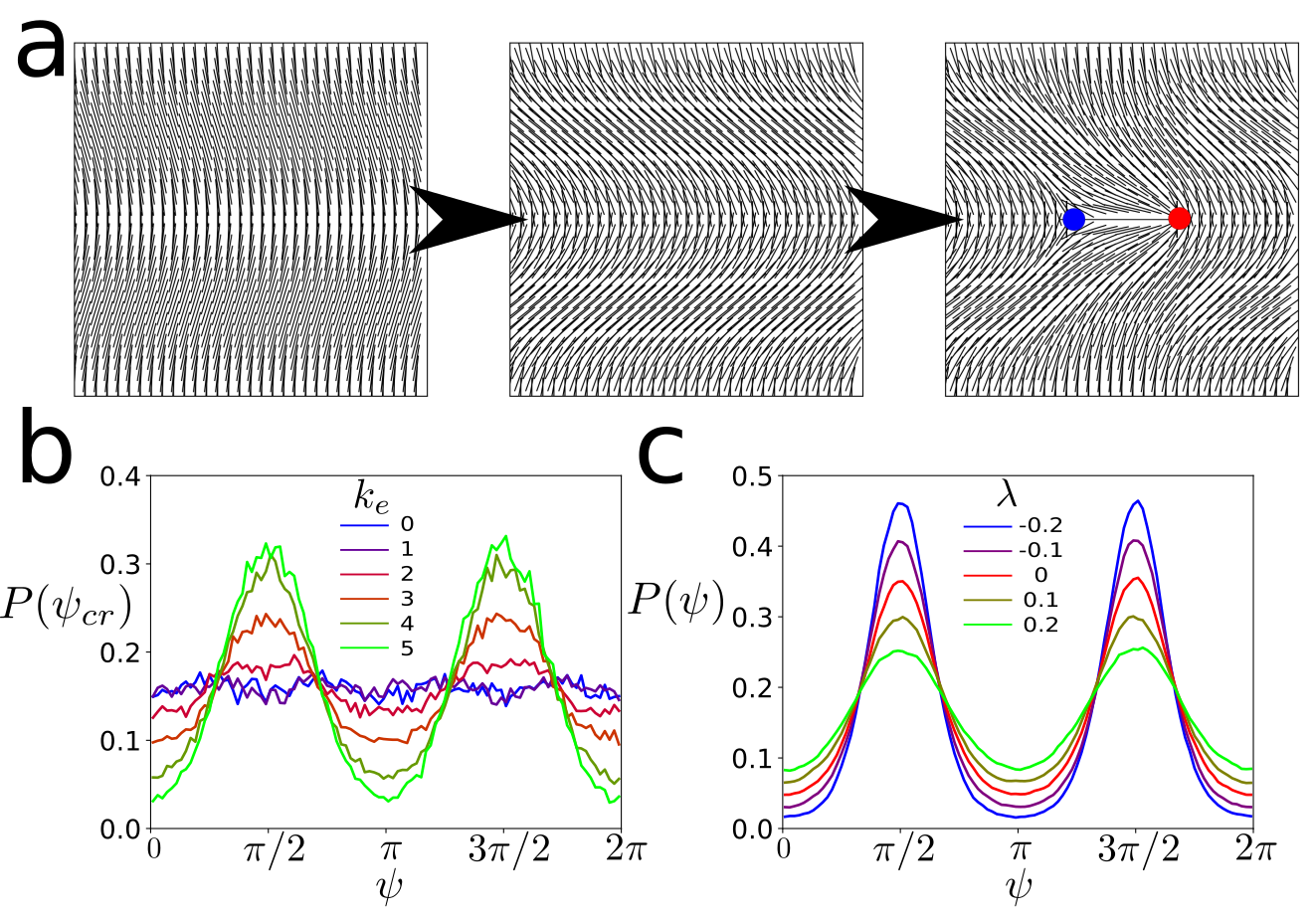}
\caption{\label{fig:mechanism} (a) Proposed mechanism. Defect free areas of the nematic tend to align with the $\hat{z}$ direction. For a extensile active nematic, these regions are unstable to bend deformations, causing defect generation perpendicular to the nematic alignment. (b) In an extensile active nematic, defects are created aligned on average perpendicular to the $\hat{z}$ direction. (c) The alignment of the defects with the cylinder depends on the flow alignment parameter, $\lambda$.}
\end{figure}

The principal difference between the passive liquid crystals and active nematics in the turbulent regime is the generation of topological defects. When the active stress is large, the ordered nematic state becomes unstable and we see defect proliferation and the so called `active turbulence' regime \cite{Giomi:2015}. The nature of this instability depends primarily on the sign of the active stress, $\alpha$ \cite{Giomi:2013,Giomi:2014}. Simple periodic deformations of the nematic director can be written in the form $\Delta\zeta = \textrm{cos}(2\pi k_ir_i)$, where $\bf{k}$ is the wave vector of the deformation and $\bf{r}$ is the position within the nematic texture. For an extensile system, $\alpha<0$, the system is unstable to bend deformations; this is when $\bf{k}$ is aligned with the nematic director ($k_i = kn_i$), see Fig.~\ref{fig:mechanism}a. When this perturbation become unstable it leads to the generation of a pair of defects aligned perpendicular to the original nematic field, see Fig.~\ref{fig:mechanism}a. When $K_e$ is increased, the nematic aligns with the $\hat{z}$ direction, hence $k_i = k\bm{\hat{e}}_z$ and the bend deformations will be of the form $\Delta\zeta = \textrm{cos}(2\pi k z)$, leading to the generation of a defect pair aligned in the $\pm\hat{\theta}$ direction. This is confirmed by looking at the orientation of defects immediately after they are created, see Fig.~\ref{fig:mechanism}b. The stability of the ordered nematic also depends on the value of the flow alignment parameter, $\lambda$ \cite{Giomi:2013,Giomi:2014,Giomi:2015,Thijssena:2020}. When $\lambda$ is increased the bend deformation is no longer the only unstable mode that can lead to defect proliferation in extensile active nematics, hence we observe a decrease in the global defect order, see Fig.~\ref{fig:mechanism}c. This mechanism relies on defect generation, explaining why increasing the activity beyond the defect proliferation threshold does not affect defect order, see Fig.~\ref{fig:active}e. Further increasing the activity slightly decreases the effect since an increased defect density leads to less order in the nematic texture, hence less regions aligned with the $\hat{z}$ direction. 

\begin{figure}[h]
	\centering
	\includegraphics[width=\columnwidth]{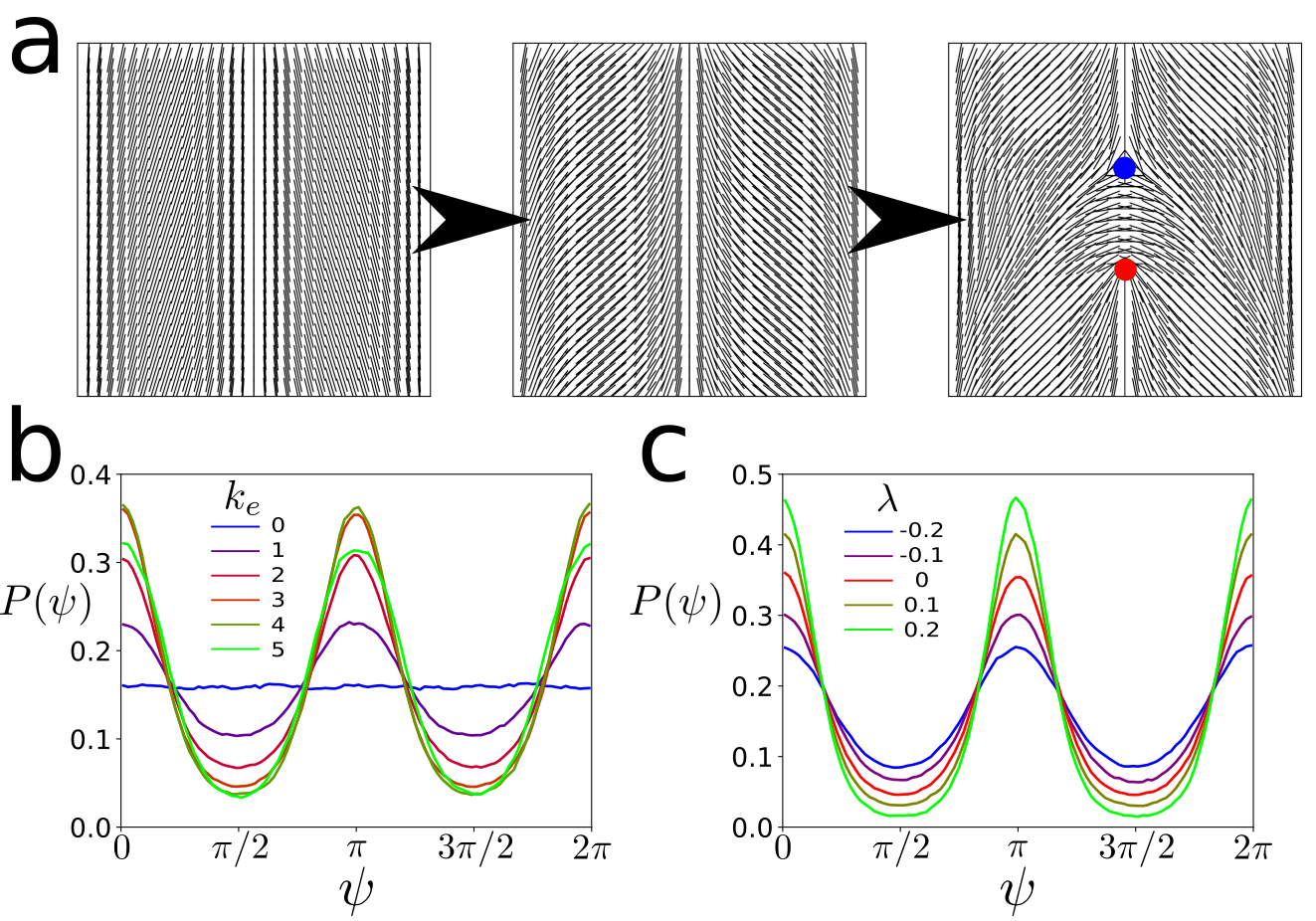}
	\caption{\label{fig:con} (a) Proposed mechanism for a contractile nematic. Ordered regions of a contractile active nematic are unstable to splay deformations, causing defect generation parallel or anti-parallel to the nematic alignment. (b) For a contractile system, extrinsic curvature coupling causes positive defects to align parallel or anti-parallel to the $\hat{z}$ direction. (c) The effect of the flow alignment parameter, $\lambda$, is reversed relative to the extensile active nematic.}
\end{figure}

This hypothesis relies on the fact that extensile active nematics are predominantly unstable to bend deformations. If the sign of the active stress is changed, $\alpha>0$, the active nematic becomes contractile in nature and the system is now predominantly unstable to splay deformations \cite{Giomi:2013,Giomi:2014}. This is when the wave vector for the perturbation is perpendicular to the nematic director ($k_i = kn_j\epsilon_{ij}$). When activity is sufficiently high, the splay deformation is unstable and leads to the generation of a pair of defects oriented parallel to the initial nematic director, see Fig.~\ref{fig:con}a. As before, the extrinsic curvature aligns the nematic director with the $\hat{z}$ direction, meaning the wave vector is now pointing in the $\hat{\theta}$ direction and splay deformations are of the form $\Delta\zeta = \textrm{cos}(2\pi k \theta)$. Therefore, when the system is contractile, $\alpha>0$, we see that the extrinsic curvature coupling causes an increased number of defects aligned perpendicular to the $\hat{z}$ direction, see Fig.~\ref{fig:con}b. Again, this anisotropy is observed in the defects at the moment of generation, see SI. In the contractile case the response to the flow alignment parameter is reversed and the effect is accentuated when $\lambda$ is increased, see Fig.~\ref{fig:con}c. This is because when $\lambda$ is decreased, the bend deformation becomes relatively less stable and defects are generated at other orientations. We observe a $\pi/2$ rotation of the nematic defect order depending on the sign of the activity, with contractile systems having defects predominantly aligned with the $\hat{z}$ direction and extensile systems having defects predominantly aligned with the $\hat{\theta}$ direction. 

By placing an active nematic on a simple curved surface it is possible to generate nematic defect order, even though the curvature does not affect the core energy of the defect. The curved surface causes the nematic to align with the smallest principle curvature, resulting in a preferred orientation for the director field. Defects are generated in pairs aligned perpendicular or parallel to the nematic if they come from bend or splay deformations, respectively. The instability of bend or splay deformations depends primarily on the nature of the active stress, with extensile active nematics being unstable to bend deformations and contractile active nematics being unstable to splay deformations. The flow alignment parameter affects the stability of the bend and splay modes differently and can be used to accentuate this phenomenon.

The mechanism outlined here can be extended to more general geometries such as a torus \cite{Ellis:2018,PearceB:2019} or textured sheets, and can be used to predict how active nematics would behave on such manifolds. In addition it can be used to measure the extrinsic curvature coupling coefficient of experimentally realized active nematics. On the surface of a cylinder the extrinsic curvature coupling is analogous to alignment to an external field or anisotropic substrate. Indeed a global defect ordered state has been observed experimentally in active nematics on an anisotropic substrate and may be explained by the effects described here combined with those caused by anisotropic hydrodynamics \cite{Guillamat:2017, Pearce:2019, Thijssena:2020}. Finally, extrinsic curvature coupling could play an important role in the organization of the cytoskeleton. This is the network of biopolymers that both gives the cell many of its mechanical properties and is able to generation motion of the cell and its components. It has already been suggested that extrinsic curvature coupling plays an important role in the location of the FtsZ division machinery at the center of a dividing bacterial cell \cite{Bisson:2017,Caldas:2019,Gonzalez:2014}. 

\section{Supplementary Information}

	\subsection{Simulation details}
	
	All results presented are generated from simulations performed on a $256\times 256$ grid. The grid spacing is set to $dx = 0.1$ and the length of a time step is set to $dt = 1\times 10^{-4}$. Equations 1-2 (main text) are solved iteratively using a stream function methodology. The stream function was solved using a Gauss Sidel algorithm accelerated using a multi-grid approach on GPU architecture. All simulation and analysis were written and performed by the author using codes written in CUDA, C++ and Python.
	
	Unless otherwise stated, the parameters for active nematic simulations were set to $\alpha = -25$, $\lambda = 0$, $\gamma = 10$, $K = 1$, $\epsilon^2 = 0.1$, $\mu_0 = 0$, $\nu_0 = 1$. For contractile active nematics, we set $\alpha = 25$. For passive nematic simulations we set $\alpha = 0$ and $\epsilon^2 = 1$.
		
	\subsection{Negative defect order}
	
	All observations hold for negative defects. The additional rotational symmetry makes the effects slightly less pronounced and have a six fold symmetry rather than two.
	
	\begin{figure}[t]
		\centering
		\includegraphics[width=0.8\columnwidth]{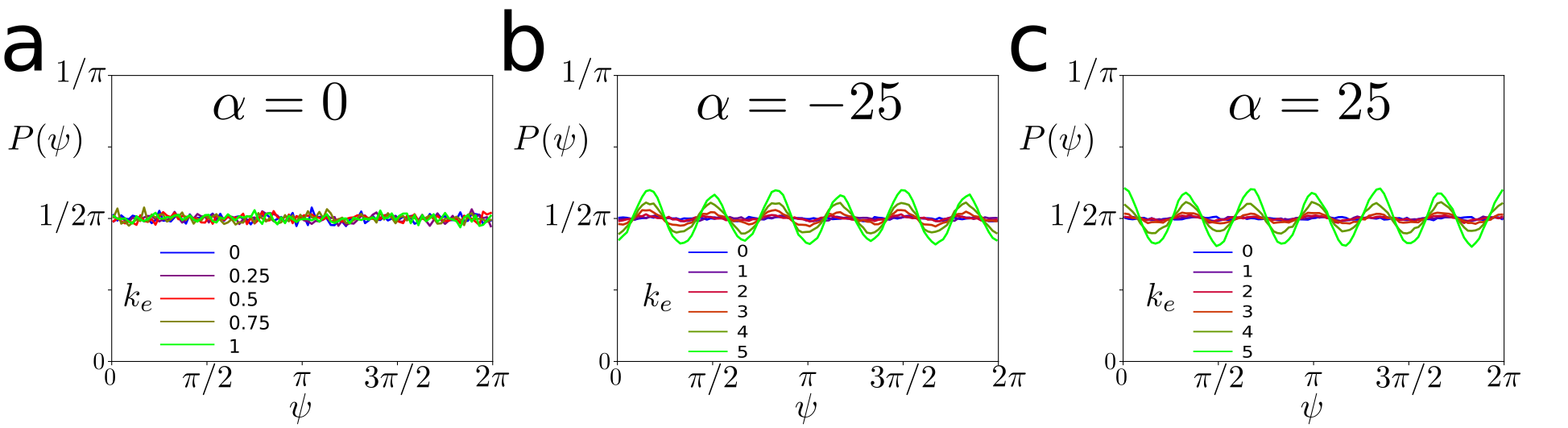}
		\caption{\label{fig:neg} The probability density function for the orientation of negative defects within an nematic on a cylindrical surface. (a) Passive nematic, complimentary to Fig.2e(Main text). (b) Active, extensile nematic ($\alpha = -25$) complimentary to Fig.1d(Main text). (c) Active, contractile nematic ($\alpha = 25$) complimentary to Fig.4b(Man text)}
	\end{figure}
	
	\subsection{Defect creation}
	
	The bend and splay instabilities that occur in active nematics are primarily dependent on the active stress. When the active stress is sufficiently strong the system becomes unstable to bend or splay depending on whether the system is extensile or contractile, respectively. Additionally, the instability depends on the flow alignment parameter, $\lambda$. For an extensile system, $\alpha < 0$, if the flow alignment parameter is large and positive both the bend and splay modes become unstable/ conversely, for a contractile system, the bend mode becomes unstable when the flow alignment parameter is large and negative. 
	
	As the extrinsic curvature coupling is increased, the anisotropy in defect generation is increased, see Fig.~\ref{fig:defcreation}a,b. The orientation of the defects as they are generated matches that predicted by the mechanism presented in the main text for both extensile and contractile active nematics, see Fig.~3,4 (Main text). Again this is accentuated for values of $\lambda$ with the same sign as $\alpha$, see Fig.~\ref{fig:defcreation}c,d.
	
	\begin{figure}[t]
	\centering
	\includegraphics[width=0.8\columnwidth]{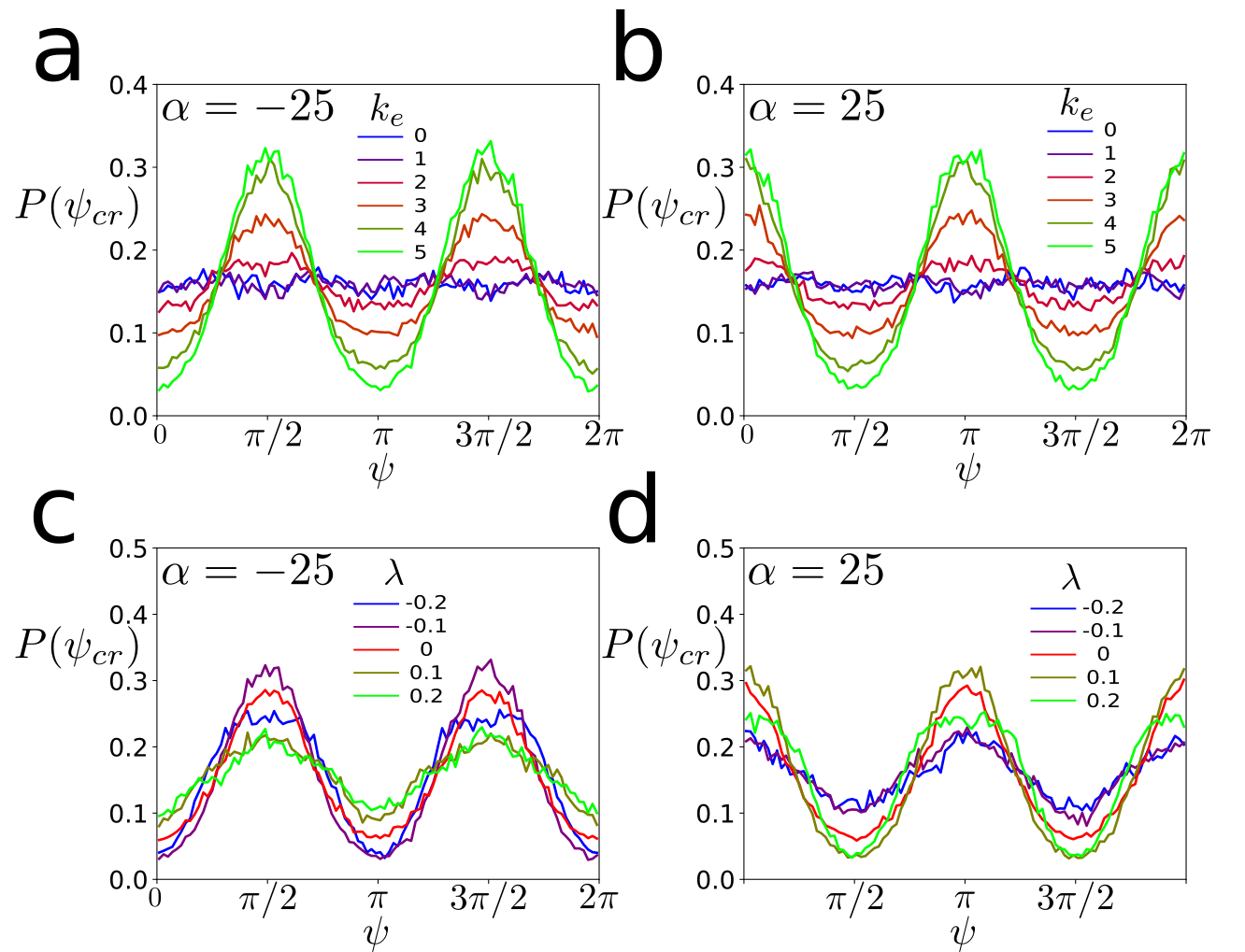}
	\caption{\label{fig:defcreation} Histogram of the orientation of defects as they are created for extensile (a,c) and contractile (b,d) active nematics. The ordering of defect creation depends on both the extrinsic curvature (a,b) and the flow alignment parameter (c,d).}
	\end{figure}

	\subsection{Effect on defect density}
	
	The active length scale is given by $l_\alpha^2 \sim K/\alpha$, which means the defect density should scale as $\rho_D\sim\alpha/K$. However in our system there is an additional source of elastic energy, associated with the extrinsic curvature coupling. If the extrinsic curvature coupling coefficient is sufficiently high, it will suppress all variations in $Q$ and lead to a defect free texture. However when $K_e$ is low, it can promote additional defect generation, as it creates well aligned regions that can then become unstable to the deformations outlined in the main text. 
	
	\begin{figure}[t]
	\centering
	\includegraphics[width=0.5\columnwidth]{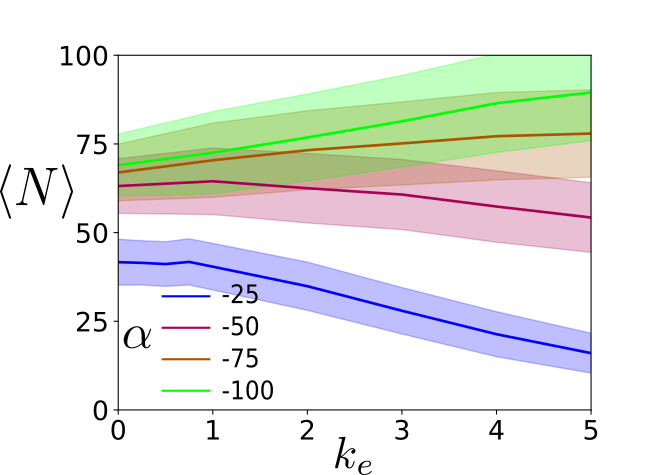}
	\caption{\label{fig:defcreation} Defect density depends primarily on the activity. When $K_e$ is sufficiently large it can suppress defect generation.}
	\end{figure}

\begin{acknowledgments}
This work was funded by the NCCR for Chemical Biology and the SNSF.
\end{acknowledgments}

\end{document}